\documentclass[
 reprint,
superscriptaddress,
 amsmath,
 amssymb,
 aps,
 prl,
longbibliography,
]{revtex4-2}
\usepackage{times}
\usepackage{amsmath}
\usepackage{amsfonts}
\usepackage{balance}
\usepackage{graphicx}
\usepackage{epstopdf}
\usepackage{dcolumn}
\usepackage{bm}
\usepackage[mathlines]{lineno}
\usepackage{amsfonts,amssymb}
\DeclareMathSizes{10}{10}{6}{4}

\begin{document}


\title{Statistical Localization of Electromagnetic Signals in Disordered Time-Varying Cavity}

\author{Bo Zhou}
\affiliation{State Key Laboratory of Extreme Photonics and Instrumentation, Zhejiang University, Hangzhou 310027, China}
\affiliation{International Joint Innovation Center, The Electromagnetics Academy at Zhejiang University, Zhejiang University, Haining 314400, China}
\author{Xingsong Feng}
\affiliation{Department of Electrical and Computer Engineering, University of California at Los Angeles, Los Angeles, CA 90095, USA}
\author{Xianmin Guo}
\affiliation{Information Materials and Intelligent Sensing Laboratory of Anhui Province, Anhui University, Hefei 230601, China}
\author{Fei Gao}
\affiliation{State Key Laboratory of Extreme Photonics and Instrumentation, Zhejiang University, Hangzhou 310027, China}
\affiliation{International Joint Innovation Center, The Electromagnetics Academy at Zhejiang University, Zhejiang University, Haining 314400, China}
\author{Hongsheng Chen}\email{corresponding authors: zuojiawang@zju.edu.cn; hansomchen@zju.edu.cn}
\affiliation{State Key Laboratory of Extreme Photonics and Instrumentation, Zhejiang University, Hangzhou 310027, China}
\affiliation{International Joint Innovation Center, The Electromagnetics Academy at Zhejiang University, Zhejiang University, Haining 314400, China}
\author{Zuojia Wang}\email{corresponding authors: zuojiawang@zju.edu.cn; hansomchen@zju.edu.cn}
\affiliation{State Key Laboratory of Extreme Photonics and Instrumentation, Zhejiang University, Hangzhou 310027, China}
\affiliation{International Joint Innovation Center, The Electromagnetics Academy at Zhejiang University, Zhejiang University, Haining 314400, China}

\begin{abstract}
In this letter, we investigate the statistical properties of electromagnetic signals after different times of duration within one-dimensional local-disordered time-varying cavities, where both spatial and temporal disorders are added. Our findings reveal that, in the vast majority of cases, adequate temporal disorder in local space can make the electromagnetic field statistically localized, obeying a normal distribution at a specific point in time of arbitrary location within the cavity. We employ the concept of disordered space-time crystals and leverage Lindeberg’s and Lyapunov’s theorems to theoretically prove the normal distribution of the field values. Furthermore, we find that with the increase of energy provided by time variation, the probability of extreme fields will significantly increase and the field intensity eventually is de-normalized, that is, deviating from the normal distribution. This study not only sheds light on the statistical properties of transient signals in local-disordered time-varying systems but also paves the way for further exploration in wave dynamics of analogous systems.
\end{abstract}

\maketitle
Wave propagation within spatially disordered systems exhibits remarkable peculiarities, challenging conventional understandings of wave dynamics \cite{RevModPhys.95.045003, RevModPhys.89.015005}. Among these peculiarities, Anderson localization stands out as a seminal phenomenon where disorder in a medium can halt the transport of waves—effectively localizing them \cite{PhysRevLett.53.2169, RevModPhys.80.1355, SegevSilberberg-650}. This phenomenon, originally proposed for electrons \cite{PhysRev.109.1492}, has since been extensively validated across various wave forms including electromagnetic \cite{SegevSilberberg-650, RevModPhys.95.045003, RevModPhys.89.015005, PhysRevLett.53.2169}, acoustic \cite{HuStrybulevych-668}, quantum \cite{CrespiOsellame-666, BillyJosse-667}, and hydrodynamic \cite{DegueldreMetzger-669}. Traditionally, the study of wave dynamics, governed by equations comprising both spatial and temporal derivatives, assumed that Anderson localization necessitated time-invariance of spatial disorder. This is changed by localization emerges from temporal discorded systems described by the Schrödinger equation \cite{PhysRevLett.119.230404, PhysRevA.94.023633, PhysRevB.96.140201}, which operates with a first-order time derivative. However, the realm of wave equations endowed with second-order temporal derivatives, such as the Maxwell's equations, remained largely unexplored until recent advancements \cite{SharabiLustig-645, CarminatiChen-646, ApffelWildeman-647, KimLee-674}. This is made possible by recent forays into the phenomena of photonic time crystals \cite{Reyes-AyonaHalevi-665, PhysRevLett.130.093803, Lustig:18, doi:10.1126/sciadv.adg7541} and space-time metamaterials \cite{CalozDeck-Léger-537, EmanueleRomain-523, doi:10.1126/science.aat3100,PhysRevLett.132.263802, ZhuZhao-699}, which have unveiled the potential for time reflections \cite{PhysRevLett.124.043902} within time-varying media. The exploration of wave dynamics within such systems reveals a rich tapestry of behaviors influenced by both the spatial and temporal characteristics of the medium.\par
Here, we build a statistical model for the local-disordered time-varying cavity (LTC) where both spatial and temporal disorder are added. Then the statistical localization properties of transient fields at specific times in one-dimensional LTCs are investigated. In the vast majority of cases, adequate temporal disorder in local space can make the field values (or electromagnetic signals) normalized at an arbitrary point within the cavity (normalization process). This phenomenon is elucidated through periodically extending of LTC alongside Lindeberg’s and Lyapunov’s Theorems, under the finite energy assumptions. Thus the LTC can be seen as disordered space-time crystals (DSTC). Additionally, we hypothesize that as the electromagnetic energy provided by the time variation increases, the probability of the emergence of extreme field values significantly increases and then causes the deviations from the normal distribution (de-normalization process). This hypothesis is confirmed by subsequent simulations. The statistical characteristics of wave fields in LTCs, particularly their tendency towards or deviates a normal distribution, can illuminate novel aspects of wave dynamics governed by time.\par
In previous work \cite{SharabiLustig-645, CarminatiChen-646}, unbounded disordered time-varying media or systems have been studied, where wave energy exhibits log-normal distribution. Since the systems are unbounded and uniform in the wave propagation direction and the time modulation of media are also needed in whole space, measuring the electromagnetic energy of the whole space is challenging. The unbounded time modulation also means that there is only temporal disorder, not spatial one. Therefore, we want to know what statistical properties the field value of a certain point will have in a smaller space—LTC, where both spatial and temporal disorders are added. We consider the one-dimensional (only z-direction) LTC as shown in Fig. 1(a), where ${L_0} = {L_1} + {L_2} + {L_3}$ is the length of the cavity, ${L_3}$ is the length of time-varying region (green), and ${L_1}$ and ${L_2}$ are the lengths of the remaining free space (white), respectively. The ends of the cavity are perfect electric conductors (PEC). Here, for simplicity, we let the permeability of the varying region $\mu  = {\mu _0}$, where ${\mu _0}$ is the permeability of free space. The permittivity of the varying region (similar with \cite{CarminatiChen-646}, plotted in Fig. 1(b)) is
\begin{equation}
    \varepsilon \left( t \right)={{\varepsilon }_{0}}\sum\nolimits_{k=0}^{\infty }{\left[ 1+{{A}_{k}}\cdot {{f}_{H}}\left( t-{{T}_{u,k}}-k{{T}_{m}} \right) \right]}\label{e1},
\end{equation}
\noindent where \({\varepsilon _0}\) is the permittivity of free space, ${A_k}$ and ${T_{u,k}}$ (\(k = 1,2, \cdots \)) are two sequences of independent identically distributed rand variables with probability density function (PDF) ${{f}_{A}}:\mathbb{R}\to [0,\infty )$ and ${{f}_{T}}:\mathbb{R}\to [0,\infty )$, respectively. ${{f}_{H}}:\mathbb{R}\to \{0,1\}$ is the Heaviside step function, and ${T_m}$ is pseudo period of time-varying.\par
\begin{figure}[htbp]
    \includegraphics{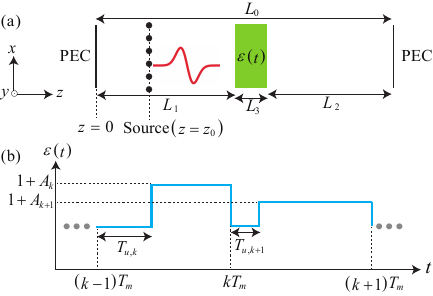}
    \caption{\label{f1}(a) Schematic of one-dimensional LTC, red line is the Gaussian pulse propagating in the LTC and green zone is the time-varying region; (b) Time-dependent permittivity of the varying region.}
\end{figure}
After the source in $z={{z}_{0}}$ emits unidirectional ($+z$) or bidirectional ($\pm z$) Gaussian pules (x-polarized, central frequency is ${{f}_{0}}={1}/{{{T}_{0}}}\;$, pulse width is ${{T}_{s}}={{N}_{s}}{{T}_{0}}$ \cite{550}), the permittivity of time-varying region start to experience random fluctuations as Eq.~(\ref{e1}). As mentioned above, the statistical properties of the field value of a certain point are what attract us. Therefore, we observe the statistical distribution of the electromagnetic signals of a certain point ${{E}_{x}}(z={{z}_{1}},t={{N}_{s}}{{T}_{0}}+{{N}_{l}}{{T}_{0}})$ after time ${{N}_{l}}{{T}_{0}}$ (${{N}_{l}}>{{N}_{s}}$). In random disordered systems, it is necessary to use a large ensemble of disorder realizations to study the statistical feature of some value \cite{SharabiLustig-645, SchwartzBartal-648}. So, a modified Finite-Difference Time-Domain (FDTD) method \cite{550} is introduced to help us to analyze the statistical feature of ${{E}_{x}}$ (MATLAB code can be found in \cite{651}) and all statistical simulations in this letter are independently repeated 5,000 times.\par
\begin{figure}[htbp]
    \includegraphics{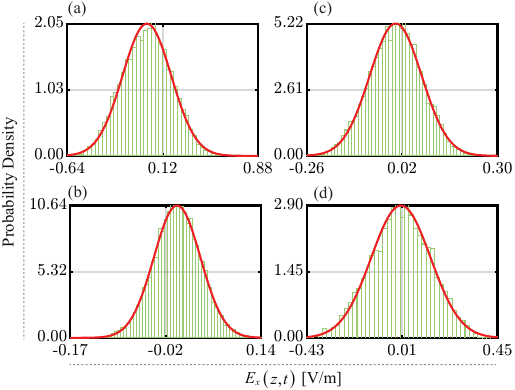}
    \caption{\label{f2}Statistical distribution of field values ${{E}_{x}}(z,t)$, where green is the histogram of the probability density and red is the corresponding normal distribution curve. Statistical distribution of (a) ${{E}_{x}}(z={10{{\lambda}_{0}}}/{128}\;,t=111.125{{T}_{0}})$, (b) ${{E}_{x}}(z={10{{\lambda}_{0}}}/{128}\;,t=201{{T}_{0}})$, (c) ${{E}_{x}}(z={10{{\lambda }_{0}}}/{128}\;,t=51{{T}_{0}})$, and (d) ${{E}_{x}}(z={10{{\lambda }_{0}}}/{128}\;,t=51{{T}_{0}})$, where (c) is results from FDTD, and (d) is results from COMSOL. ${{\lambda }_{0}}={{{c}_{0}}}/{{{f}_{0}}}\;$ is the wavelength and ${{c}_{0}}={1}/{\sqrt{{{\mu }_{0}}{{\varepsilon }_{0}}}}\;$ is the speed of light in free space. For specific parameter settings, see  \cite{550}.
}
\end{figure}
For simplicity, we let ${{A}_{k}}$ and ${{T}_{u,k}}$ obey uniform distribution in $[{{A}_{lb}},{{A}_{ub}}]$ and $[{{T}_{lb}},{{T}_{ub}}]$ \cite{550}. We begin by letting ${{A}_{lb}}=0$, ${{A}_{ub}}=1$, ${{T}_{lb}}=0.2$, and ${{T}_{ub}}=0.8$. Then we find that after certain  period of time, the field values ${{E}_{x}}$ at almost all points in the LTC follow a normal distribution, as shown in Fig. 2(a)-(c) (additional figures can be found in \cite{550}). To ensure the accuracy of the modified FDTD, we also run the simulations using time-domain solver of COMSOL \cite{550} under the same settings and obtain similar results to FDTD, as shown in Fig. 2(d). In these LTCs, it seems that the electromagnetic signals always obey a normal distribution, after a certain period of random perturbation in the $\varepsilon (t)$. This is independent of the direction of Gaussian pulse, the pulse width of Gaussian pulse, position of the varying region, and the length of the LTC.\par
This intriguing phenomenon is similar to the Anderson localization that occurs on spatially disordered systems where some intensity logarithm of the system always obeys a normal distribution \cite{SharabiLustig-645, DeychLisyansky-649, SegevSilberberg-650}. Since wave energy also exhibits log-normal distribution in the discorded time crystal \cite{SharabiLustig-645, CarminatiChen-646}, to explain this phenomenon, we convert the LTC into a DSTC. By removing the PEC boundaries and periodically extending the one-dimensional LTC in the z-direction, we obtain the one-dimensional DSTC consisting of $\text{2}N+1$ identical LTCs with the PEC boundaries removed, as shown in Fig. 3(a)-(b). Without the PEC, the Gaussian pulse is free to propagate outward. Since the wave is decelerated in the time-varying region, the wave reaches as far as the $N\text{-th}$ extended cavity ($N=\left\lceil {({{N}_{s}}+{{N}_{l}}){{\lambda }_{0}}}/{{{L}_{0}}}\; \right\rceil $, $\left\lceil \cdot  \right\rceil $ denotes rounding up) in $t=({{N}_{s}}+{{N}_{l}}){{T}_{0}}$. The $\text{2}N+1$ LTCs are identical, thus the field value of LTC ${{E}_{x}}(z,t)$ can be calculated by the field value of DSTC $E_{x}^{D}(z,t)$ as
\begin{align}
   {{E}_{x}}(z,t)=&\sum\nolimits_{k=-\left\lceil N/2 \right\rceil }^{\left\lceil N/2 \right\rceil }{\left[ R_{0}^{2k}E_{x}^{D}(z+2k{{L}_{0}},t) \right.} \nonumber\\ 
 & \left. +R_{0}^{2k-1}E_{x}^{D}(-z+2k{{L}_{0}},t) \right]  \label{e2},
\end{align}
where ${{R}_{0}}=-1$ is the reflection coefficient of PEC. \par
\begin{figure*}[htbp]
    \includegraphics{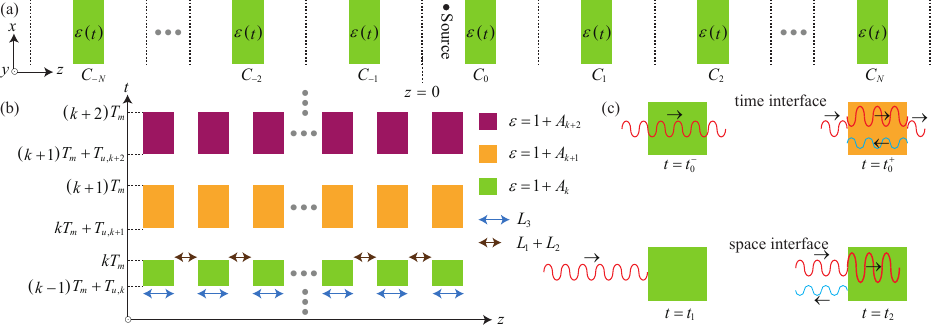}
    \caption{\label{f3}(a) Schematic of one-dimensional DSTC. Green is the time-varying region with same $\varepsilon (t)$ and the source is still only one at $z={z}_{0}$; (b) Space-time diagram of a one-dimensional DSTC. Different colors represent different magnitudes of relative permittivity. In the spatial dimension, they are periodically extended, so the widths are both ${{L}_{3}}$ and the intervals are both ${{L}_{1}}+{{L}_{2}}$; but in the time dimension, they are randomly disordered, so the intervals and widths are not the same. (c) Schematic of time interface and space interface, where red lines denote forward or transmitted wave and blue lines denote backward or reflected wave. 
}
\end{figure*}
For the Gaussian pulse in the DSTC, it will experience multiple time interfaces (fluctuations in permittivity occurring in the time-varying regions, e.g., $\varepsilon (z,t=t_{0}^{-})={{\varepsilon }_{1}}{{\varepsilon }_{0}}$ and $\varepsilon (z,t=t_{0}^{+})={{\varepsilon }_{2}}{{\varepsilon }_{0}}$) and space interfaces (spatial discontinuities of permittivity, e.g., $\varepsilon (z=z_{0}^{-},t)={{\varepsilon }_{1}}{{\varepsilon }_{0}}$ and $\varepsilon (z=z_{0}^{+},t)={{\varepsilon }_{2}}{{\varepsilon }_{0}}$), as shown in Fig. 3(c). The time interface forces the wave (amplitude ${{E}_{0}}$, wave vector ${{\mathbf{k}}_{0}}$ and frequency ${{f}_{0}}$) in the time-varying region to split into forward (${{\mathsf{\mathcal{T}}}_{t}}{{E}_{0}}$, ${{\mathbf{k}}_{0}}$ and $\sqrt{{{{\varepsilon }_{1}}}/{{{\varepsilon }_{2}}}\;}{{f}_{0}}$) and backward (${{\mathsf{\mathcal{R}}}_{t}}{{E}_{0}}$, ${{\mathbf{k}}_{0}}$ and $-\sqrt{{{{\varepsilon }_{1}}}/{{{\varepsilon }_{2}}}\;}{{f}_{0}}$) wave \cite{CalozDeck-Léger-537, EmanueleRomain-523}, where ${{\mathsf{\mathcal{T}}}_{t}}={{{\varepsilon }_{1}}(1+\sqrt{{{{\varepsilon }_{2}}}/{{{\varepsilon }_{1}}}\;})}/{2{{\varepsilon }_{2}}}\;$ and ${{\mathsf{\mathcal{R}}}_{t}}={{{\varepsilon }_{1}}(1-\sqrt{{{{\varepsilon }_{2}}}/{{{\varepsilon }_{1}}}\;})}/{2{{\varepsilon }_{2}}}\;$, which corresponds to temporal disorder. The space interface forces the wave to split into transmitted (${{\mathsf{\mathcal{T}}}_{s}}{{E}_{0}}$, ${{\mathbf{k}}_{0}}$ and ${{f}_{0}}$) and reflected (${{\mathsf{\mathcal{R}}}_{s}}{{E}_{0}}$, $-{{\mathbf{k}}_{0}}$ and ${{f}_{0}}$) wave \cite{Kong-519}, where ${{\mathsf{\mathcal{T}}}_{s}}={2}/{(1+\sqrt{{{{\varepsilon }_{2}}}/{{{\varepsilon }_{1}}}\;})}\;$ and ${{\mathsf{\mathcal{R}}}_{s}}={(1-\sqrt{{{{\varepsilon }_{2}}}/{{{\varepsilon }_{1}}}\;})}/{(1+\sqrt{{{{\varepsilon }_{2}}}/{{{\varepsilon }_{1}}}\;})}\;$, which corresponds to spatial disorder. In our model, there are two-type time and space interfaces: the first is {${{\varepsilon }_{1}}=1$, ${{\varepsilon }_{2}}=1+{{A}_{k}}$} with $\mathsf{\mathcal{T}}_{t12}^{k}$, $\mathsf{\mathcal{R}}_{t12}^{k}$, $\mathsf{\mathcal{T}}_{s12}^{k}$ and $\mathsf{\mathcal{R}}_{s12}^{k}$; the second is {${{\varepsilon }_{1}}=1+{{A}_{k}}$, ${{\varepsilon }_{2}}=1$} with $\mathsf{\mathcal{T}}_{t21}^{k}$, $\mathsf{\mathcal{R}}_{t21}^{k}$, $\mathsf{\mathcal{T}}_{s21}^{k}$ and $\mathsf{\mathcal{R}}_{s21}^{k}$ (note that all $\mathsf{\mathcal{R}}$ and $\mathsf{\mathcal{T}}$ are random variables). Then the field value of DSTC can be written as

\begin{equation}
    E_{x}^{D}(z,t)=\sum\nolimits_{j}{E_{x,j}^{D}\cos {{\Theta }_{j}}}
    \label{e3},
\end{equation}
\begin{equation}
    E_{x,j}^D = \prod\limits_{{k_i} \in \alpha _i^j} {\left( {\mathcal{T}_{s12}^{{k_1}}\mathcal{R}_{s12}^{{k_2}}\mathcal{T}_{s21}^{{k_3}}\mathcal{R}_{s21}^{{k_4}}} \right)} \prod\limits_{{k_i} \in \beta _i^j} {\left( {\mathcal{T}_{t12}^{{k_1}}\mathcal{R}_{t12}^{{k_2}}\mathcal{T}_{t21}^{{k_3}}\mathcal{R}_{t21}^{{k_4}}} \right)}
    \label{e4},
\end{equation}
where $\alpha _{i}^{j}$ and $\beta _{i}^{j}$ ($i=1,2,3,4$) are some random subsets of $\{x|x\le N,x\in \mathbb{N}\}$ and related to $(z,t)$. ${{\Theta }_{j}}$ (called “cumulative phase” \cite{CarminatiChen-646}) is also a random variable and related to $(z,t)$. Obtaining statistical properties of $\alpha _{i}^{j}$, $\beta _{i}^{j}$ and ${{\Theta }_{j}}$ is difficult, since the time interfaces only work for waves that are inside the time-varying region, but not for those outside. Besides, it is an intractable problem that $\alpha _{i}^{j}$, $\beta _{i}^{j}$ and ${{\Theta }_{j}}$ are related to $(z,t)$. However, we can try to ignore the effect of the time interfaces to make some estimates of $E_{x}^{D}(z,t)$, since only a part of the region is time-varying in DSTC and ${{T}_{m}}>{{T}_{0}}$. Temporal interface will provide energy and cause the field value to increase exponentially, but not space interface \cite{CalozDeck-Léger-537, EmanueleRomain-523,Kong-519,PanCohen-652}. In practical time-modulated systems, there is inevitably an energy loss, and if this loss is comparable to the gain due to the time modulation, then the energy of the whole system will converge \cite{ApffelWildeman-647}. Without time interfaces, energy of the entire DSTC is conserved. Pulse width ${{T}_{s}}={{N}_{s}}{{T}_{0}}$ is finite and there is no any singular point, thus the energy of DSTC is finite. Therefore, we can write $\max \left| E_{x}^{D}(z,t) \right|={{C}_{s}}<+\infty $ even in lim ${{N}_{l}}\to \infty $ (which means $N\to \infty $), where $t\in [0,({{N}_{s}}+{{N}_{l}}){{T}_{0}}]$ and $z\in [-({{N}_{s}}+{{N}_{l}}){{\lambda }_{0}},({{N}_{s}}+{{N}_{l}}){{\lambda }_{0}}]$.\par
Similar with \cite{CarminatiChen-646}, we consider that $E_{x}^{D}(z+2k{{L}_{0}},t)$ and $E_{x}^{D}(-z+2k{{L}_{0}}-{{L}_{0}},t)$ are independent random variables. For convenience, we rewrite Eq.~(\ref{e2}) as
\begin{equation}
    {{E}_{x}}(z,t)={{E}_{N}}=\sum\nolimits_{k=-\left\lceil N/2 \right\rceil }^{\left\lceil N/2 \right\rceil }{{{X}_{k}}}
    \label{e5},
\end{equation}
where ${{X}_{k}}=R_{0}^{2k}E_{x}^{D}(z+2k{{L}_{0}},t)+R_{0}^{2k-1}E_{x}^{D}(-z+2k{{L}_{0}},t)$ are independent random variables. With this condition, it is natural to use the Lindeberg’s Theorem or the Lyapunov’s Theorem \cite{Ash-654,Billingsley-653,Loève-657,Grimmett-656} to prove that ${{E}_{x}}(z,t)$ is converging to a normal distribution, as shown in Fig. 2. Here, we give a characterization of Lyapunov’s Theorem: Let ${{S}_{n}}=\sum _{k=1}^{n}{{Y}_{k}}$, where ${{Y}_{k}}$ are independent random variables with finite mean ${{m}_{k}}$ and finite variance $\sigma _{k}^{2}$. Let ${{T}_{n}}=c_{n}^{-1}({{S}_{n}}-\mathbb{E}[{{S}_{n}}])$, where $c_{n}^{2}=\text{Var(}{{\text{S}}_{n}})=\sum _{k=1}^{n}\sigma _{k}^{2}$. Assume that for some $\delta >0$
\begin{equation}
    \underset{n\to \infty }{\mathop{\lim }}\,\frac{1}{c_{n}^{2+\delta }}\sum\nolimits_{k=1}^{n}{\mathsf{\mathbb{E}}\left[ {{\left| {{Y}_{k}}-{{m}_{k}} \right|}^{2+\delta }} \right]}=0
    \label{e6},
\end{equation}
then ${{T}_{n}}$ converges in distribution to a random variable that is normal with mean 0 and variance 1, denoted by ${{T}_{n}}\xrightarrow{d}N(0,1)$. Eq.~(\ref{e6}) is called Lyapunov’s condition.\par
It can be proved that mean of ${{X}_{k}}$ (still denoted by ${{m}_{k}}$) and variance of ${{X}_{k}}$ (still denoted by $\sigma _{k}^{2}$) are finite \cite{550}, since $\left| {{X}_{k}} \right|<2{{C}_{s}}<\infty $. We assume that $c_{N}^{2}=\sum _{k=-\left\lceil N/2 \right\rceil }^{\left\lceil N/2 \right\rceil }\sigma _{k}^{2}\to \infty $ as lim $N\to \infty $, which does not contradict $\sigma _{k}^{2}<\infty $ \cite{zb1}. Let ${{Z}_{N}}=c_{N}^{-1}({{E}_{N}}-\mathbb{E}[{{E}_{N}}])$, ${{X}^{'}_{k}}={{X}_{k}}-{{m}_{k}}$, ${{E}^{'}_{N}}=\sum _{k=-\left\lceil N/2 \right\rceil }^{\left\lceil N/2 \right\rceil }{{X}^{'}_{k}}$, ${c}^{'2}_{N}=\text{Var(}{{E}^{'}_{N}}\text{)}$ and ${{Z}^{'}_{N}}={c}_{N}^{'-1}({{E}^{'}_{N}}-\mathbb{E}[{{E}^{'}_{N}}])$, then we have ${{m}^{'}_{k}}=\mathrm{}[{{X}^{'}_{k}}]=0$, ${\sigma }_{k}^{'2}=\sigma _{k}^{2}$, ${c}_{N}^{'2}=c_{N}^{2}$ and $\left| {{X}^{'}_{k}} \right|\le \left| 2{{C}_{s}}-{{m}_{k}} \right|<\infty $. With these conditions, it can be proved that Lyapunov condition is satisfied for $\delta =1$ and thus ${{Z}^{'}_{N}}\xrightarrow{d}N(0,1)$ \cite{550, Ash-654,Billingsley-653,Loève-657,Grimmett-656}. When the Lyapunov’s Theorem holds, then Lindeberg’s Theorem holds, since the former is a sufficient condition for the latter \cite{550, Ash-654,Billingsley-653}. Then we can prove that Lindeberg’s Theorem holds for ${{E}_{N}}$, since Lindeberg’s Theorem holds for ${{E}^{'}_{N}}$ \cite{550, Ash-654}. Therefore, we obtain ${{Z}_{N}}\xrightarrow{d}N(0,1)$, which means that ${{E}_{x}}(z,t)={{E}_{N}}$ converges in distribution to random variable that is normal with mean ${{M}_{N}}=\sum _{k=-\left\lceil N/2 \right\rceil }^{\left\lceil N/2 \right\rceil }{{m}_{k}}$ and variance $c_{N}^{2}$. \par
Now we have proved that ${{E}_{x}}(z,t)$ converges in distribution to normal distribution as $t\to \infty $ ($N\to \infty $), which means that the larger $t$ is, the closer the statistical distribution of ${{E}_{x}}(z,t)$ is to a normal distribution (normalization process). However, this is the result of ignoring the energy gain from the time interface. If we consider the LTC to be lossless, the energy gain provided by the time interface is also small and negligible when $t$ is small; however, when $t$ is sufficiently large, the energy gain cannot be ignored, and thus the probability of ${{E}_{x}}(z,t)$ appearing to be extremely large increases significantly (see additional figures in \cite{550}). Then, we can speculate that as $t$ increases, the distribution of ${{E}_{x}}(z,t)$ should converge to a normal distribution and then deviate, that is, be normalized and then de-normalized. To verify our conjecture, we introduce the Cramér-von Mises goodness-of-fit tests for normality \cite{Stephens-664,Anderson-661,CsörgőFaraway-663, 10.1111/j.2517-6161.1980.tb01100.x}. The tests are implemented in the R language \cite{R}. Null hypothesis is set as ${{E}_{x}}(z,t)$ is normal distribution, $p$ is the probability of observing a test statistic as extreme as, or more extreme than, the observed value under the null hypothesis. Small values of $p$ cast doubt on the validity of the null hypothesis, which means the larger $p$ is, the more ${{E}_{x}}(z,t)$ follows a normal distribution. \par
In Fig. 4(a), for all three cases shown, $p$-values show an increase and then a decrease, which implies that ${{E}_{x}}(z,t)$ does indeed converge to a normal distribution and then deviate, as predicted. Since the percentage of time-varying regions is smaller in Case 2 and Case 3 (50\%) than in Case 1 (25\%), the time-varying provides more energy, so the $p$-values of Case 2 and Case 3 decrease earlier ($1000<2000$). Similarly, we examine the effect of the range of variation in permittivity $[{{A}_{lb}},{{A}_{ub}}]$ on the results, since the larger the variation is, the more energy is provided. In Fig. 4(b), for all three cases shown, the $p$-values show a decrease when ${{A}_{ub}}$ is large enough and ${{A}_{lb}}=0$, which implies that ${{E}_{x}}(z,t)$ does indeed deviate from a normal distribution since more energy is provided. Again, due to the percentage of time-varying regions, the $p$-values of Case 2 and Case 3 decrease earlier ($4<8$).\par
\begin{figure}[htbp]
    \includegraphics{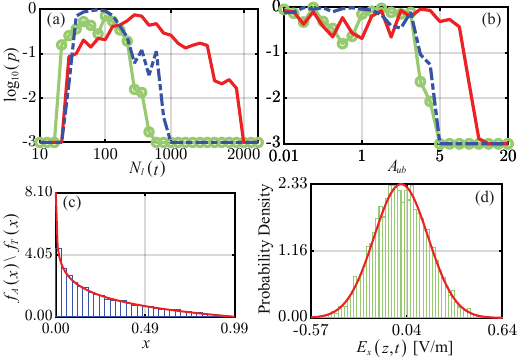}
    \caption{\label{f4}(a) $p$-value of ${{E}_{x}}(z,t)$ , fixed ${{A}_{ub}}=1$ and varying ${{N}_{l}}$ from 10 to 1000. (b) $p$-value of ${{E}_{x}}(z,t)$, fixed ${{N}_{l}}=100$ and varying ${{A}_{ub}}$ from 0.01 to 20. In (a) and (b), red line denotes the Case 1 where $({{L}_{1}},{{L}_{2}},{{L}_{3}})=({3{{\lambda }_{0}}}/{8}\;,{3{{\lambda }_{0}}}/{8}\;,{{{\lambda }_{0}}}/{4}\;)$, green line with circle denotes the Case 2 where $({{L}_{1}},{{L}_{2}},{{L}_{3}})=({{{\lambda }_{0}}}/{8}\;,{{{\lambda }_{0}}}/{8}\;,{{{\lambda }_{0}}}/{4}\;)$ and blue dashed line denotes the Case 3 where $({{L}_{1}},{{L}_{2}},{{L}_{3}})=({{{\lambda }_{0}}}/{4}\;,{{{\lambda }_{0}}}/{4}\;,{{{\lambda }_{0}}}/{2}\;)$. When the $p$-value is less than 0.001, i.e., $\log_{10}(p)<-3$, the probability that the null hypothesis holds is extremely small. Therefore, for ease of graphical presentation, truncation is done for even smaller $p$-value, all of which are shown in the graph as $\log_{10}(p)=-3$. (c) PDF of a non-uniform distribution of ${{A}_{k}}$ and ${{T}_{u,k}}$, and (d) its corresponding statistical distribution of ${{E}_{x}}(z={10{{\lambda }_{0}}}/{128}\;,t=201{{T}_{0}})$ when $({{L}_{1}},{{L}_{2}},{{L}_{3}})=({{{\lambda }_{0}}}/{4}\;,{{{\lambda }_{0}}}/{4}\;,{{{\lambda }_{0}}}/{2}\;)$. For all these cases, ${{N}_{s}}=1$, ${{z}_{0}}={4{{\lambda }_{0}}}/{128}\;$, ${{T}_{m}}=10{{T}_{0}}$, and Gaussian pules are unidirectional. 
}
\end{figure}
Recall our proof process that we do not assume the specific forms of ${{f}_{A}}$ and ${{f}_{T}}$. Therefore, it is possible to observe a normal distribution of ${{E}_{x}}(z,t)$ for any ${{f}_{A}}$ and ${{f}_{T}}$ that can satisfy $\max \left| E_{x}^{D}(z,t) \right|={{C}_{s}}<+\infty $. For example, we let ${{f}_{A}}(x)={{f}_{T}}(x)=-\ln x$, $0<x<1$, as shown in Fig. 4(c). We still observe a normal distribution of ${{E}_{x}}(z,t)$ when $({{L}_{1}},{{L}_{2}},{{L}_{3}})=({{{\lambda }_{0}}}/{4}\;,{{{\lambda }_{0}}}/{4}\;,{{{\lambda }_{0}}}/{2}\;)$, as shown in Fig. 4(d). This means that the normal distribution of ${{E}_{x}}(z,t)$ is insensitive to ${{f}_{A}}$ and ${{f}_{T}}$. This has obvious implications for future experimental measurements. Besides, we have not resorted to the properties of the wave equation, so these findings are generalizable to other systems governed by the wave equation, such as water waves, sound waves, and so on. This also reduces the difficulty of experimental observations. Similarly, since ${{T}_{m}}$ can be much larger than ${{T}_{0}}$, this reduces the difficulty of realizing a time modulation system, which can be conveniently implemented in electromagnetic wave systems in most frequency bands. For example, in radio frequency regime, dynamic transmission line \cite{Reyes-AyonaHalevi-665, HuidobroGaliffi-697} is a good choice. As shown in Fig. 4(b), even ${{A}_{ub}}=0.01$, a norm distribution can be still obtained, so the modulation depth is no longer an issue. Besides, we consider the unavoidable losses in real systems, the normal distribution can still be observed after adding certain losses in the simulation (see additional figures in \cite{550}). All these bring great convenience for our future experimental observation.\par
To summarize, we explore the novel idea of disordered time-varying cavities, focusing on the statistical characteristics of electromagnetic signals at specific points. Through FDTD and COMSOL simulations, we found these values are normally distributed. To explain this, we applied disordered space-time crystal concepts and cited Lindeberg’s and Lyapunov’s Theorems to verify our findings. We hypothesize and later confirm that increased energy alters these distributions, raising the probability of extreme values. For future prospects, the exploration of LTCs presents intriguing avenues for further research. One compelling direction involves investigating the implications of multiple time-varying regions within the cavity. How would interactions between these regions influence the statistical distribution of field values? Would correlated or uncorrelated time-varying domains lead to distinct field dynamics? Addressing these questions could deepen our understanding of complex wave interactions in disordered systems, potentially revealing emergent phenomena analogous to those observed in spatially disordered environments. Overall, this study not only paves the way for future experiments but also extends its implications to wave-dominated systems beyond electromagnetism, including  hydrodynamic and acoustics.\par
The work was sponsored by the National Natural Science Foundation of China (62222115, 62171407), the Key Research and Development Program of Zhejiang Province under Grant No.2024C01241(SD2), and the Fundamental Research Funds for the Central Universities.\par

\nocite{*}

\begin{thebibliography}{46}
\bibitem [{zb1()}]{zb1}%
  \BibitemOpen
  \href@noop {} {\bibinfo  {journal} {If $c_{N}^{2}<\infty $, then $\sigma _{k}^{2}\to 0$ and ${{X}_{k}}$ degenerates to a constant, which contradicts the assumption of random variable and is not of interest to us}\ }\BibitemShut {NoStop}%
\end{thebibliography}
%

\end{document}